\documentclass[aps,prd,superscriptaddress,showpacs,preprintnumbers]{revtex4}
\usepackage{graphicx,color,here,epsf,psfig}

\begin{document}

\newcommand{\dlt}{\bigtriangleup}
\newcommand{\beq}{\begin{equation}}
\newcommand{\eeq}[1]{\label{#1} \end{equation}}
\newcommand{\insertplot}[1]{\centerline{\psfig{figure={#1},width=14.5cm}}}

\parskip=0.3cm


\title{$J/\Psi$ PHOTOPRODUCTION IN A DUAL MODEL}

\author{R. Fiore}
\affiliation{Universita' della Calabria,
Dipt. di Fisica \\
 Arcavacata di Rende, I-87030 Cosenza, Calabria, Italy}

\author{L.L. Jenkovszky}
\affiliation{Bogolyubov Institute for Theoretical Physics (BITP),
Ukrainian National Academy of Sciences \\14-b, Metrolohichna str.,
Kiev, 03143, Ukraine}

\author{V.K.~Magas}
\affiliation{Departament d'Estructura i Constituents de la
Mat\'eria,\\ Universitat de Barcelona, Diagonal 647,\\
08028 Barcelona, Spain}

\author{F. Paccanoni}
\affiliation{Universita' di Padova, Dipartimento di Fisica and
INFN, sezione di Padova,\\ via Marzolo 8, I-35100 Padua, Italy}

\author{A.~Prokudin}
\affiliation{Dipartimento di Fisica Teorica, Universit\`a di Torino and \\
          INFN, Sezione di Torino, Via P. Giuria 1, I-10125 Torino, Italy}

\begin{abstract}
$J/\Psi$ photoproduction is studied in the framework of the
analytic $S-$matrix theory. The differential and integrated
elastic cross sections for $J/\Psi$ photoproduction are calculated
from a Dual Amplitude with Mandelstam Analyticity (DAMA). It is
argued that at low energies, the background, which is the
low-energy equivalent of the high-energy diffraction replaces the
Pomeron exchange. The onset of the high energy Pomeron dominance
is estimated from the fits to the data.\\
\\
{\it Dedicated to Professor
Anatoly I. Bugrij on the occasion of his 60-th birthday.}
\end{abstract}

\pacs{11.55.-m, 11.55.Jy, 12.40.Nn}

\maketitle

\section{Introduction} \label{s1}

$J/\Psi$ photoproduction is a unique testing field for
diffraction. Most of the theoretical approaches to $J/\Psi$
photoproduction are based on the Pomeron or multi-gluon exchanges
in the $t$-channel of the reaction (for a review see
\cite{review}). A common feature of these models is the
uncertainty of the  low-energy extrapolation of the high-energy
exchange mechanisms. The missing piece is the low-energy
background contribution, significant between the threshold and the region of the
dominance of the exchange mechanism, e.g., the onset of the
Regge-Pomeron asymptotic behaviour.

What is the fate of the Regge exchange contribution when
extrapolated to low energies? The answer to this question was
given in late 60-ies by dual models: the proper sum of
direct-channel resonances produces Regge asymptotic behavior and
vice versa. The alternative, i.e. taking the sum of the two (the
so-called interference model) is incorrect, resulting in double
counting.

While the realization of the Regge-resonance duality was
quantified within narrow-resonance dual models, notably in the
Veneziano model \cite{Veneziano}, a similar solution for the
Pomeron (=diffraction) was not possible in the framework of that
model, just because of its narrow-resonance nature. Another reason
for the poor understanding of "low-energy diffraction", or the
background, is difficulties in its separation (discrimination
and identification), because of the mixing with the resonance part.

Experimentally, the identification of these two components meets
difficulties coming from the flexibility of their
parameterizations. The separation of a Breit-Wigner resonance from
the background, as well as the discrimination of some $t-$channel
exchanges with identical flavor content and C-parity, e.g. the
Pomeron mixing with the $f$-meson or the odderon with $\omega,$ is
a familiar problem in experimental physics.

Now we have an ideal opportunity in hands: $J/\Psi$
photoproduction is purely diffractive since resonances here are
not produced and, due to the Okubo-Zweig-Iizika (OZI) rule
\cite{OZI}, no flavor (valence quarks) can be exchanged in the
$t$-channel. The derivation of the OZI rule is based on simple
quark diagrams and they are closely related to the so-called
duality quark diagrams, to be discussed in the next section.

In the next section we shortly remind the reader the basics of the
two-component duality. In Section 3 a dual model applicable to
both the diffractive and non-diffractive (resonance) components of
the amplitude is introduced. Its application to $J/\Psi$
photoproduction is presented in Section 4. The onset of the Regge
asymptotic behavior is shown in Section 5, where a number of
various models of Pomeron trajectories is also compared. A short
summary of the paper is given in Section 6.

\section{Two-Component Duality}
According to our present knowledge about two-body hadronic
reactions, two distinct classes of reaction mechanisms exist.
\begin{figure}
\begin{center}
\includegraphics[width=0.8\textwidth
]{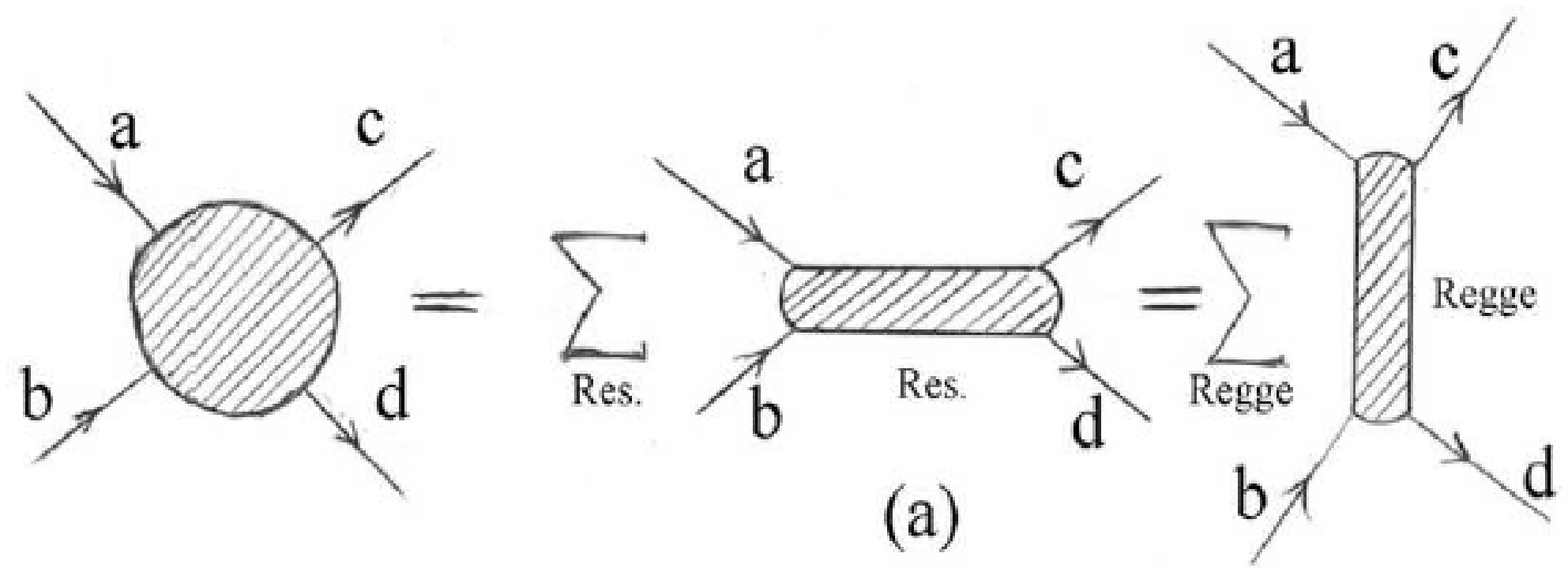}
\includegraphics[width=0.7\textwidth
]{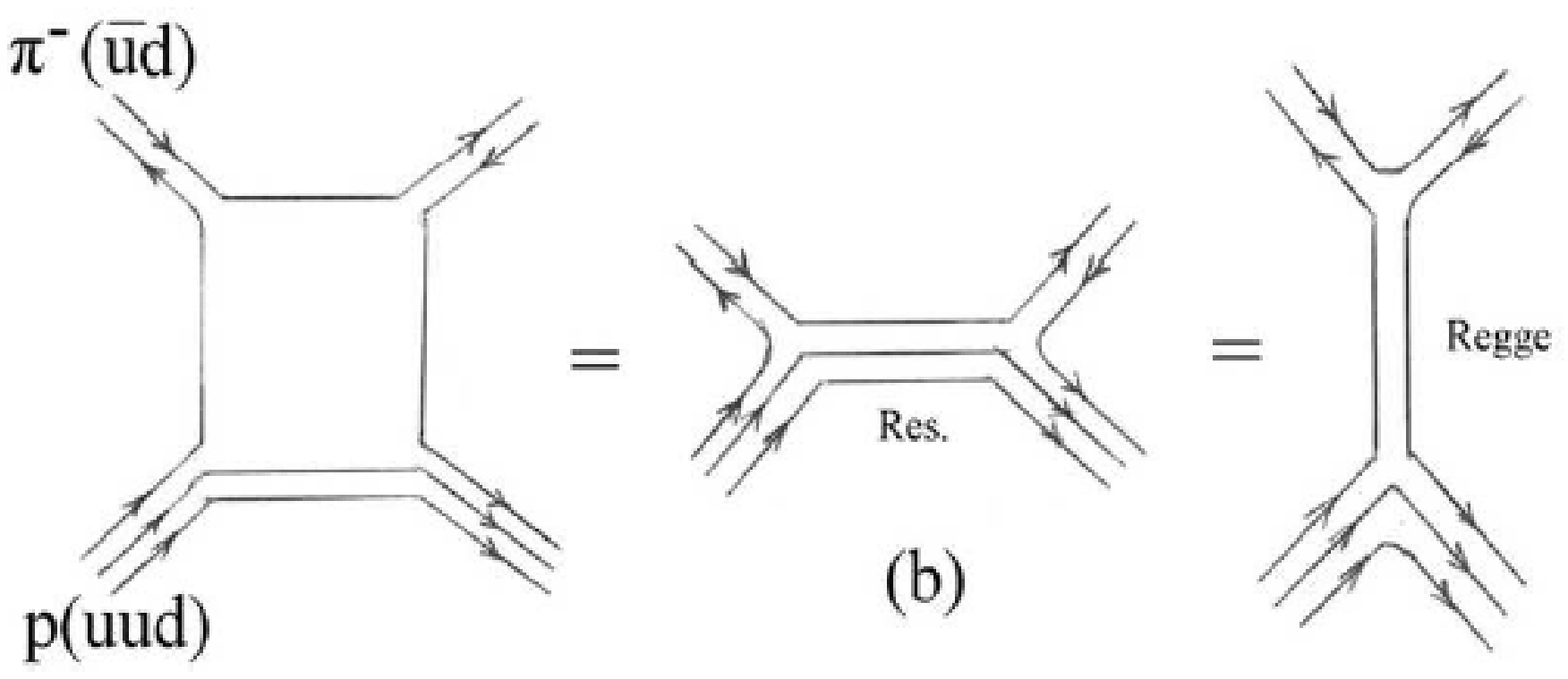}
\end{center}
\caption{\label{fig:duality} a) By duality, the proper sum of
resonances in the direct channel produces its asymptotic Regge
behavior and v.v. (upper panel); b) the same equality in terms of
duality quark diagrams (lower panel).}
\end{figure}

The first one includes the formation of resonances in the $s-$channel 
and the exchange of particles, resonances, or Regge
trajectories in the $t-$channel. The low-energy, resonance
behavior and the high-energy, Regge asymptotics are related by
duality, illustrated in Fig. 1a, which at Born level, or,
alternatively, for tree diagrams, mathematically can be formalized
in the Veneziano model, which is a combination of Euler
Beta-functions \cite{Veneziano}.

\begin{figure}
\begin{center}
\includegraphics[width=0.6\textwidth]{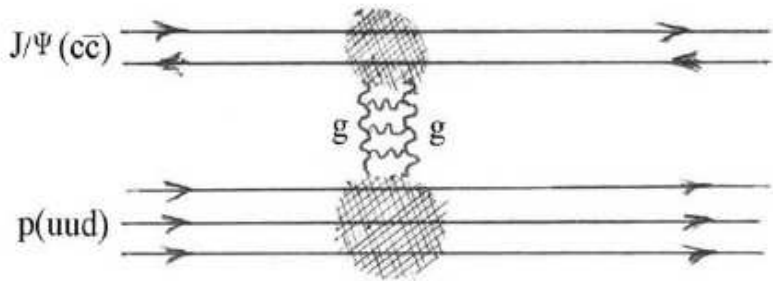}
\end{center}
\caption{\label{quark_diag}  Duality quark diagram for elastic
$J/\Psi-p$ scattering.}
\end{figure}

The second class of mechanisms does not exhibit resonances at low
energies and its high-energy behavior is governed by the exchange
of a vacuum Regge trajectory, the Pomeron, with an intercept equal
to or slightly greater than one. Harari and Rosner \cite{H_R}
hypothesized that the low-energy non-resonating background is dual
to the high-energy Pomeron exchange, or diffraction. In other
words, the low energy background should extrapolate to high-energy
diffraction in the same way as the sum of narrow resonances sum up
to produce Regge behaviour. However, contrary to the case of
narrow resonances, the Veneziano amplitude, by construction,
cannot be applied to (infinitely) broad resonances. This becomes
possible in a generalization of narrow resonance dual models
called dual amplitudes with Mandelstam analyticity (DAMA),
allowing for (infinitely) broad resonances \cite{superbroad}, or
the background.

Duality has an elegant interpretation in terms of quark-partons,
called dual quark diagrams (Figs. 1 and 2). Theses diagram can be
deformed by conserving their topology as shown in Fig. 1b, where
direct-channel resonances correspond to the quark box squeezed in
the horizontal direction and Regge exchange corresponds to
squeezing it vertically.

Resonances can be formed only in non-exotic channels, made of a
quark and antiquark pair (meson) or three quarks (baryon), as e.g.
in the $\pi^-p$ channel, shown in Fig. 1b. Resonances cannot be
produced in exotic channels, e.g. in proton-proton scattering (a
two-baryon state) or in $K^+ p$ scattering, while $p\bar p$ and
$K^- p,$ are non-exotic and exhibit observable resonances.
According to the two-component duality hypotheses \cite{H_R}, the
non-resonating direct-channel background is dual to a Pomeron
exchange. By drawing a duality quark diagram, it is easy to see
that, by the OZI rule, a quark-antiquark pair cannot be exchanged
in the $t$ channel of an exotic reaction. The interaction can be
mediated by multi-gluon (Pomeron) exchange, as illustrated in Fig.
2 for $J/\Psi-p$ scattering.

The above simple rules are confirmed experimentally: total cross
section of $p\bar p$ and $K^- p$ scattering exhibit a rich
resonance structure (non-exotic channels) and a rapid decrease at
high energies due to the decreasing contribution from sub-leading
reggeons, while $p\bar p$ and $K^+ p$ total cross sections are
nearly flat due to the non-resonant low-energy background dual to
the high-energy Pomeron exchange -- both being of diffractive
nature. The above rules are not exact due to the presence of mixed
states, violation of exchange degeneracy of $t$ channel
trajectories, unitarity corrections (non-planar diagrams) etc.
These effects are less important in $J/\Psi-p$ scattering.

The two-component dual picture of hadronic dynamics is summarized
in Table I.

\begin{table}
\caption{Two-component
duality}
\begin{tabular}{|c|c|c|}
  \hline
  ${\cal I}m A(a+b\rightarrow c+d)=$& $R$& Pomeron \\
  \hline
  $s-$channel & $\sum A_{Res}$  & Non-resonant background \\
  \hline
  $t-$channel & $\sum A_{Regge}$ & Pomeron $(I=S=B=0;\ C=+1)$ \\
  \hline
  Duality quark diagram & Fig. 1b & Fig. 2 \\
  \hline
  High energy dependence & $s^{\alpha-1},\ \alpha<1$ & $s^{\alpha-1},\ \alpha\geq 1$ \\
  \hline
\end{tabular}
\end{table}

\section{Dual Model} \label{s2}

Dual models with Mandelstam analyticity \cite{DAMA} appeared as a
generalization of narrow-resonance (e.g. Veneziano) dual models,
intended to overcome the manifestly non-unitarity of the latter.
Contrary to narrow-resonance dual models, DAMA does not only
allows for, but moreover requires  non-linear, complex
trajectories. This property allows for the presence in DAMA of
finite-width resonances and a non-vanishing imaginary part of the
amplitude. The maximal number of direct resonances is correlated
with the maximal value of the real part of the relevant trajectory
or, alternatively, with the mass of its heaviest threshold. An
extreme case is when the real part of the trajectory terminates
below the spin of the lowest ($s-$channel) resonance, called
"super-broad-resonance approximation" \cite{superbroad}, in
contrast with the narrow-resonance approximation, e.g., of the
Veneziano amplitude. The resulting scattering amplitude can
describe the non-resonating direct-channel background, dual to
the Pomeron exchange in the $t-$channel. The dual properties of
this construction were studied in Ref. \cite{Jenk}.

The $(s,t)$ term ($s$ and $t$ are Mandelstam variables) of a dual
amplitude with Mandelstam analyticity (DAMA) \cite{DAMA} is given
by: \beq D(s,t)=c \int_0^1 {dz \biggl({z \over g}
\biggr)^{-\alpha(s')-1} \biggl({1-z \over
g}\biggr)^{-\alpha_t(t')}}\,, \eeq{dama} where $\alpha(s)$ and
$\alpha(t)$ are Regge trajectories in the $s$ and $t$ channel
correspondingly; $s'=s(1-z), \ \ t'=tz$; $g$ and $c$ are
parameters, $g>1$, $c>0$.

For $s\rightarrow\infty$ and fixed $t$ DAMA, eq. (1), is
Regge-behaved
\beq
D(s,t)\sim s^{\alpha_t(t)-1}.
\eeq{ass}

In the vicinity of the threshold, $s \rightarrow s_0$, \beq D(s,t)
\sim \sqrt{s_0-s}\, [const+\ln(1-s_0/s)]\,. \eeq{threshold}

The pole structure of DAMA is similar to that of the Veneziano
model except that multiple poles appear on daughter levels
\cite{DAMA}. \beq D(s,t)=\sum_{n=0}^{\infty}
g^{n+1}\sum_{l=0}^{n}\frac{[-s\alpha'(s)]^{l}C_{n-l}(t)}
{[n-\alpha(s)]^{l+1}},\,. \eeq{series} where $C_n(t)$ is the
residue, whose form is fixed by the $t$-channel Regge trajectory
(see \cite{DAMA}) \beq
C_l(t)=\frac{1}{l!}\frac{d^l}{dz^l}\left[\biggl({1-z \over
g}\biggr)^{-\alpha_t(tz)}\right]_{z=0}\,. \eeq{p5} The pole term
in DAMA is a generalization of the Breit-Wigner formula,
comprising a whole sequence of resonances lying on a complex
trajectory $\alpha(s)$. Such a "reggeized" Breit-Wigner formula
has little practical use in the case of linear trajectories,
resulting in an infinite sequence of poles, but it becomes a
powerful tool if complex trajectories with a limited real part and
hence a restricted number of resonances are used.

A simple model of trajectories satisfying the threshold and
asymptotic constraints is a sum of square roots \cite{DAMA} \beq
\alpha(s)\sim \sum_i\gamma_i\sqrt{s_i-s}.
 \eeq{traj}
 The number of thresholds included depends on the model. While the
 lightest threshold gives the main contribution to the imaginary part,
 the heaviest one promotes the rise of the real part (terminating
 at the heavies threshold).

 A particular case of the model eq. (\ref{traj}) is that with a single
 threshold
\beq \alpha(s)=\alpha(0)+\alpha_1(\sqrt{s_0}-\sqrt{s_0-s})\, .
\eeq{trajectory} Imposing an upper bound on the real part of this
 trajectory, $Re\, \alpha(s)<0$,
 we get an
amplitude that does not produce resonances, since the real part of
the trajectory does not reach $n=0$ where the first pole could
appear. This is the ansatz we suggest for the exotic trajectory.
The imaginary part of such a trajectory instead rises
indefinitely, contributing to the total cross section with a
smooth background.

The super-broad-resonance approximation \cite{superbroad}, in a
sense, is opposite to the narrow-resonance approximation, typical
of the Veneziano model. However, contrary to the latter, valid
only when the resonances widths vanish, the super-broad resonance
approximation allows for a smooth transition to observable
resonances or to the "Veneziano limit" \cite{DAMA}. Dual
properties of this model were studied in \cite{Jenk}.

\section{$J/\Psi$ Photoproduction}
\label{ss2}

Photoproduction of vector mesons is well described in the
framework of the vector meson dominance model \cite{VMD},
according to which the photoproduction scattering amplitude $A$ is
proportional to the sum of the relevant hadronic amplitudes
\cite{Collins} (see also \cite{ACW}):
 \beq
 D_H(\gamma\, P\rightarrow V\, P)=\sum_V{e\over{f_V}}D_H(V\, P\rightarrow V\, P),
 \eeq{VMD}
where $e$ is the vector meson-photon coupling constant, $V=\rho,
\omega, \phi, J/\Psi, ...$ Within this approximation,
photoproduction is reduced to elastic hadron scattering ($J/\psi-p$
in our case), where the constants $e$ and $f_V$ are absorbed by
the normalization factor, to be fitted to the data.

Among various vector mesons we choose one, namely $J/\psi$ since,
by the OZI rule \cite{OZI}, in $J/\psi-p$ scattering only the
Pomeron trajectory can be exchanged in the $t$ channel. To a
lesser extent, this is true also for the $\phi-p$, however in the
latter case ordinary meson exchange is present due to
$\omega-\phi$ mixing. Heavier vector mesons are as good as
$J/\Psi-p$, but relevant data are less abundant. So, we find
$J/\Psi-p$ scattering to be an ideal testing field (filter) for
diffraction: in the direct channel only exotic trajectories are
allowed and they are dual to the exchange of the Pomeron
trajectory. Diffraction can be studied uncontaminated by secondary
trajectories. This possibility was already emphasized in earlier
publications, see \cite{Kononenko} and refs.
\cite{DAMA,superbroad}, however without fits to the experimental
data.

In the present paper we apply DAMA for meson-baryon scattering
\footnote{By having accepted vector dominance, we thus reduce
$J/\Psi$ photoproduction to $J/\psi-p$ scattering} with an exotic
trajectory in the direct channel and the Pomeron trajectory in the
exchange channel and calculate the differential and integrated
elastic cross section for $J/\psi-p$ photoproduction. The
parameters of the model are fitted to the experimental data on
$J/\Psi$ photoproduction. With these parameters we calculate also
the imaginary part of the forward amplitude proportional to the
$J/\psi-p$ total cross section.

In the framework of the Regge pole models, $J/\Psi$
photoproduction was studied in numerous papers. Apart from the
flexibility inherent in the Regge pole approach, there is an
ambiguity in the low energy behavior, below the Regge asymptotics.
Usually, this low-energy domain is either ignored by a lower bound
in the applications, or it is accounted for by the inclusion of a
threshold factor. While in the first case one eliminates part of
the dynamics, below the (assumed) Regge asymptotic behavior, the
second option ignores the domain between the threshold and Regge
asymptotic behavior, characterized by resonances in
non-diffractive processes, otherwise mimicked by a direct-channel
exotic contribution.

Following \cite{JKP}, we write the meson-baryon elastic scattering
amplitude (with $J/\Psi$ photoproduction in mind) as a combination
\beq 
D(s,t,u)=(s-u)\,(D(s,t) - D(u,t))\, . 
\eeq{AAA} 
For the exotic Regge trajectory such as (7) the
scattering amplitude is given by convergent integral, eq. (9)
with (1), and can be calculated for any $s$
and $t$ without analytical continuation, needed otherwise, as
discussed in \cite{DAMA}.

\begin{figure}
\begin{center}
\includegraphics[width=0.45\textwidth,bb= 10 140 540 660]{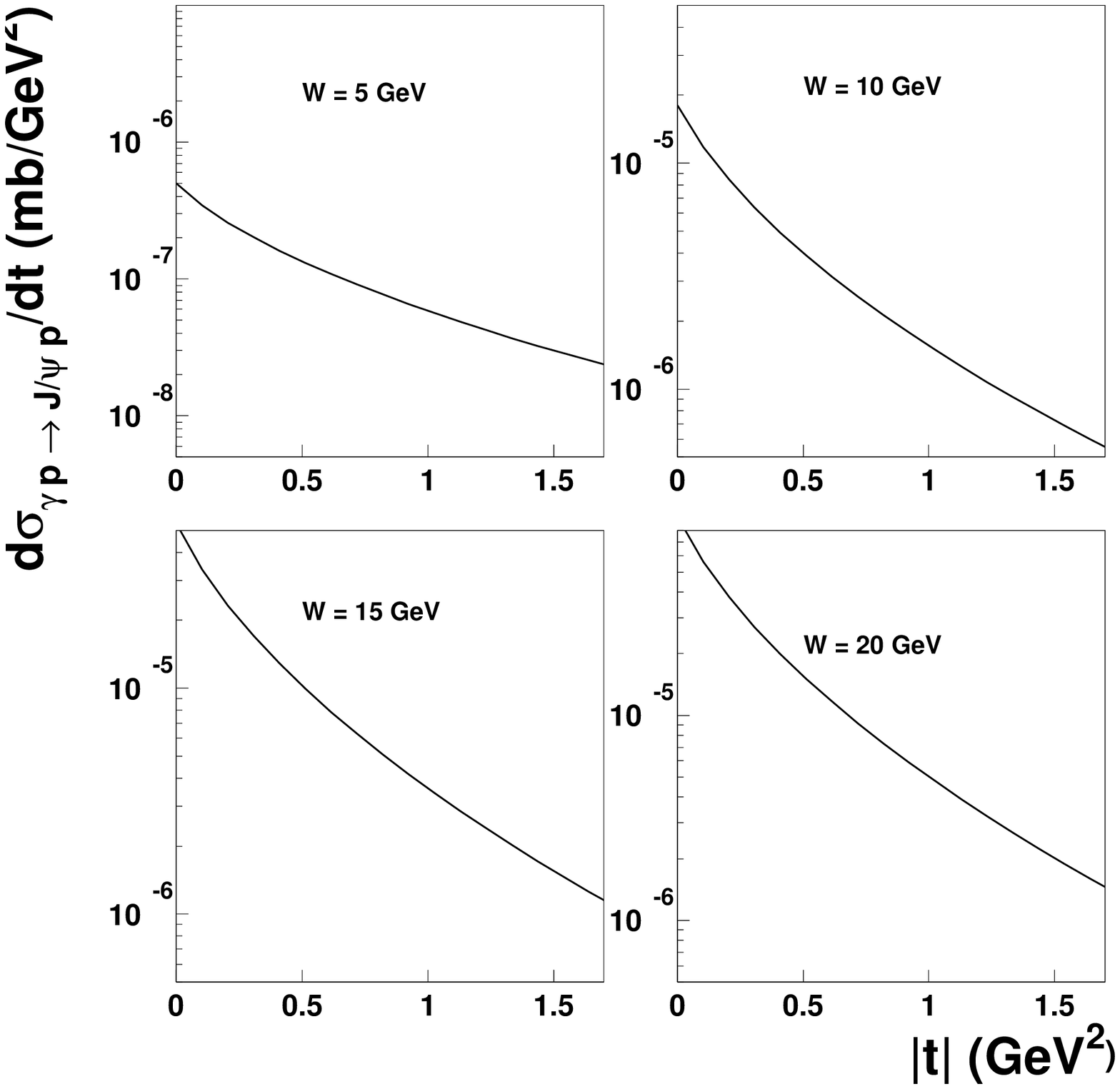}
\includegraphics[width=0.45\textwidth,bb= 10 140 540 660]{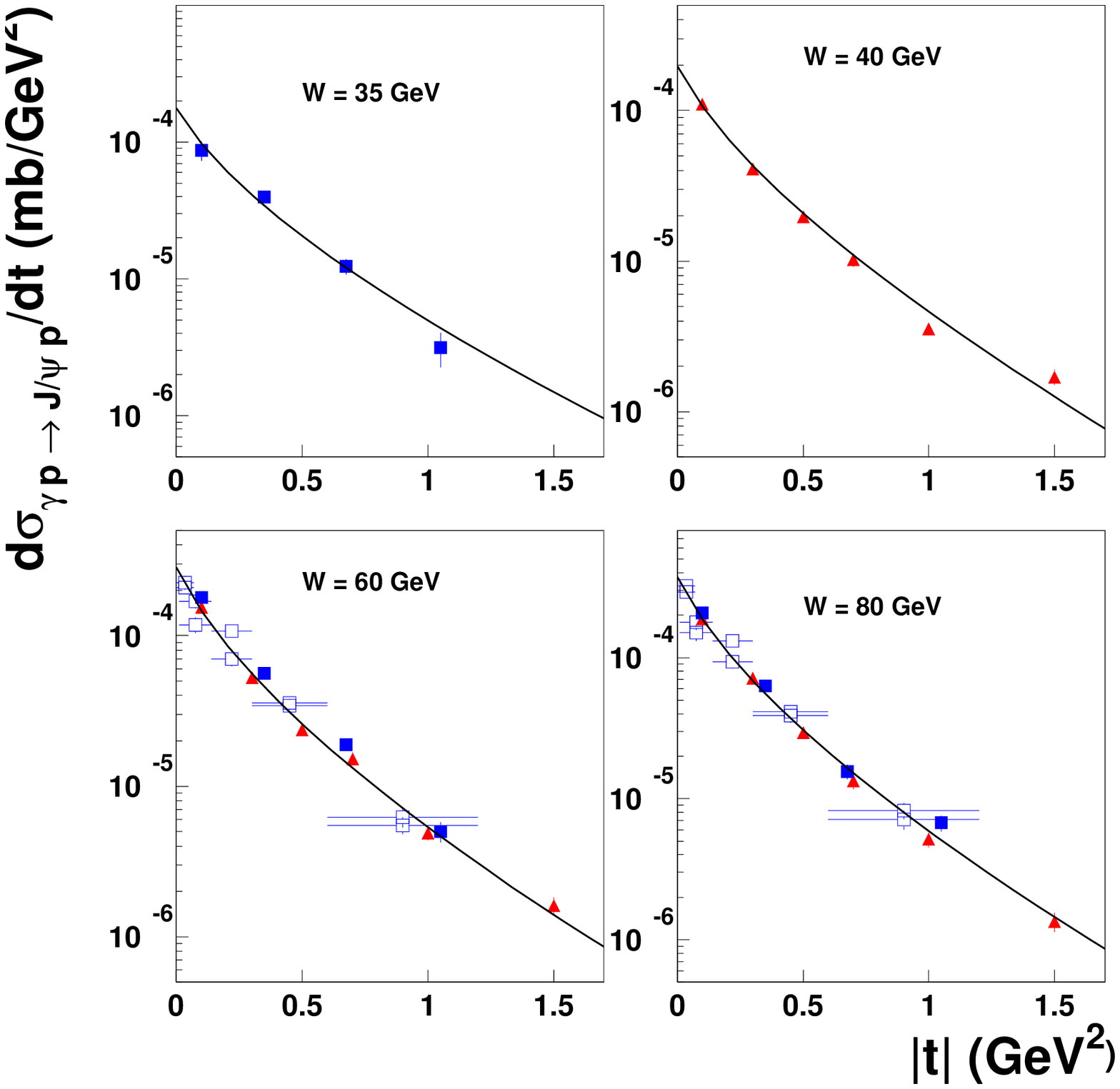}
\end{center}
\caption{\label{fig:jpsi_df}  $J/\Psi$ differential cross sections
as a function of $t$ in the energy range $W$ between 5 and 80
GeV.}
\end{figure}

We use a $t$-channel Pomeron trajectory in the form 
\beq
\alpha^P(t)=\alpha^P(0)+\alpha^P_1(\sqrt{t_1}-\sqrt{t_1-t})+
2\alpha^P_2(t_2-\sqrt{(t_2-t)t_2}) 
\eeq{Pomer} with a light
(lowest) threshold $t_1=4m_{\pi}^2$ and a heavy one $t_2$, whose
value, together with other parameters appearing in (1), will be
fitted to the data (see Table I). Eq. (\ref{Pomer}) is a generalization of 
the trajectory with one square root threshold used in \cite{FJPP}b). 
Throughout this paper
$\alpha(t)$ and $\alpha^P(t)$ are identical since channel only the
Pomeron trajectory contributes in the $t$.

Arguments from the fits in favor
of a nonlinear trajectory can be found
also in Ref. \cite{Brandt} as well as in Regge pole models
\cite{FJPP, ProkMart} fitting high-energy $J/\Psi$ photo- and
electroproduction. The behavior of these trajectories has much in
common at small (above, say $-2$ GeV$^2$, whereafter they may strongly deviate.
In Sec. 6 we compare their behavior and add more comments on that.

\begin{figure}
\begin{center}
\includegraphics[width=0.45\textwidth,bb= 10 140 540 660]{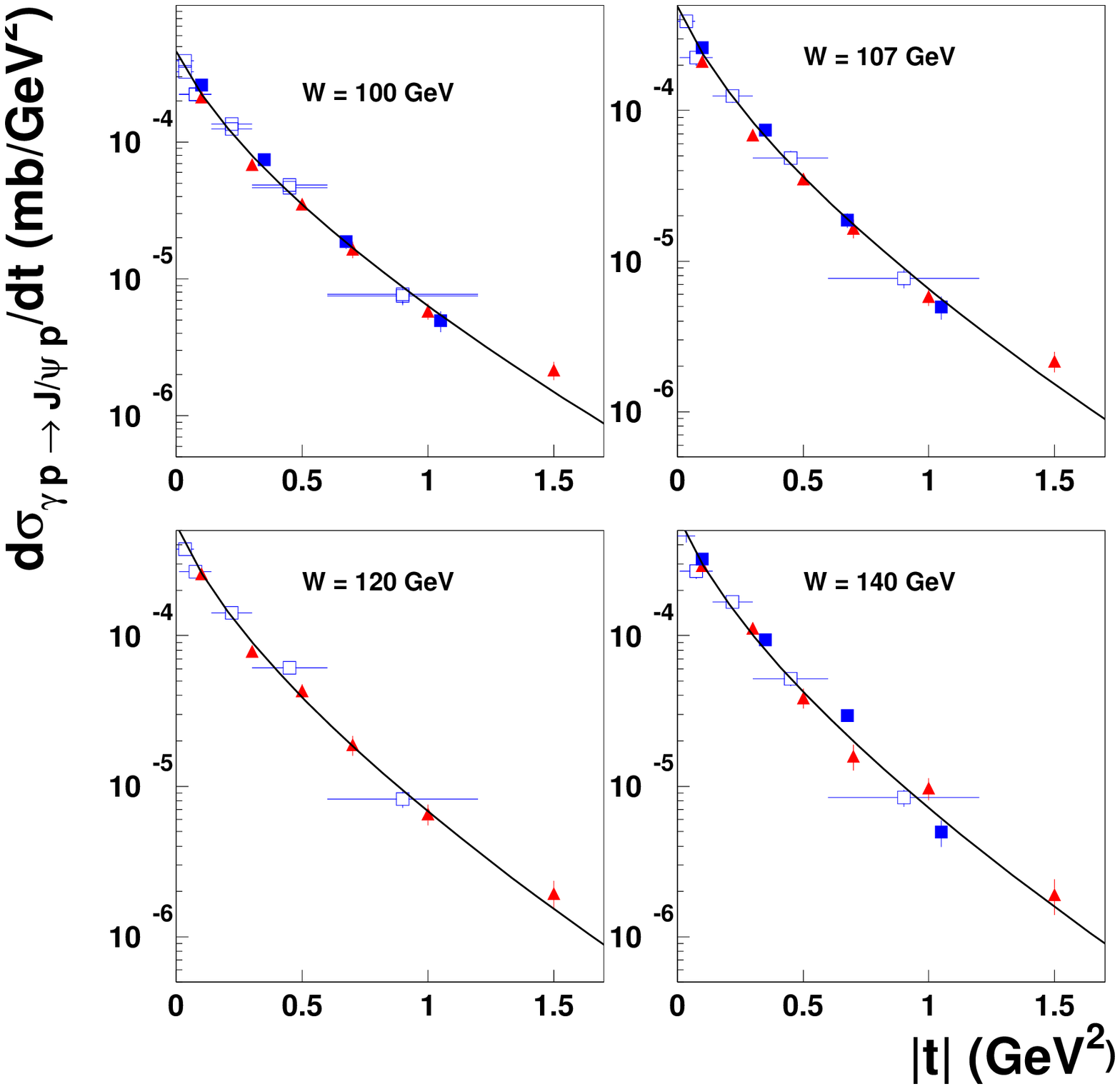}
\includegraphics[width=0.45\textwidth,bb= 10 140 540 660]{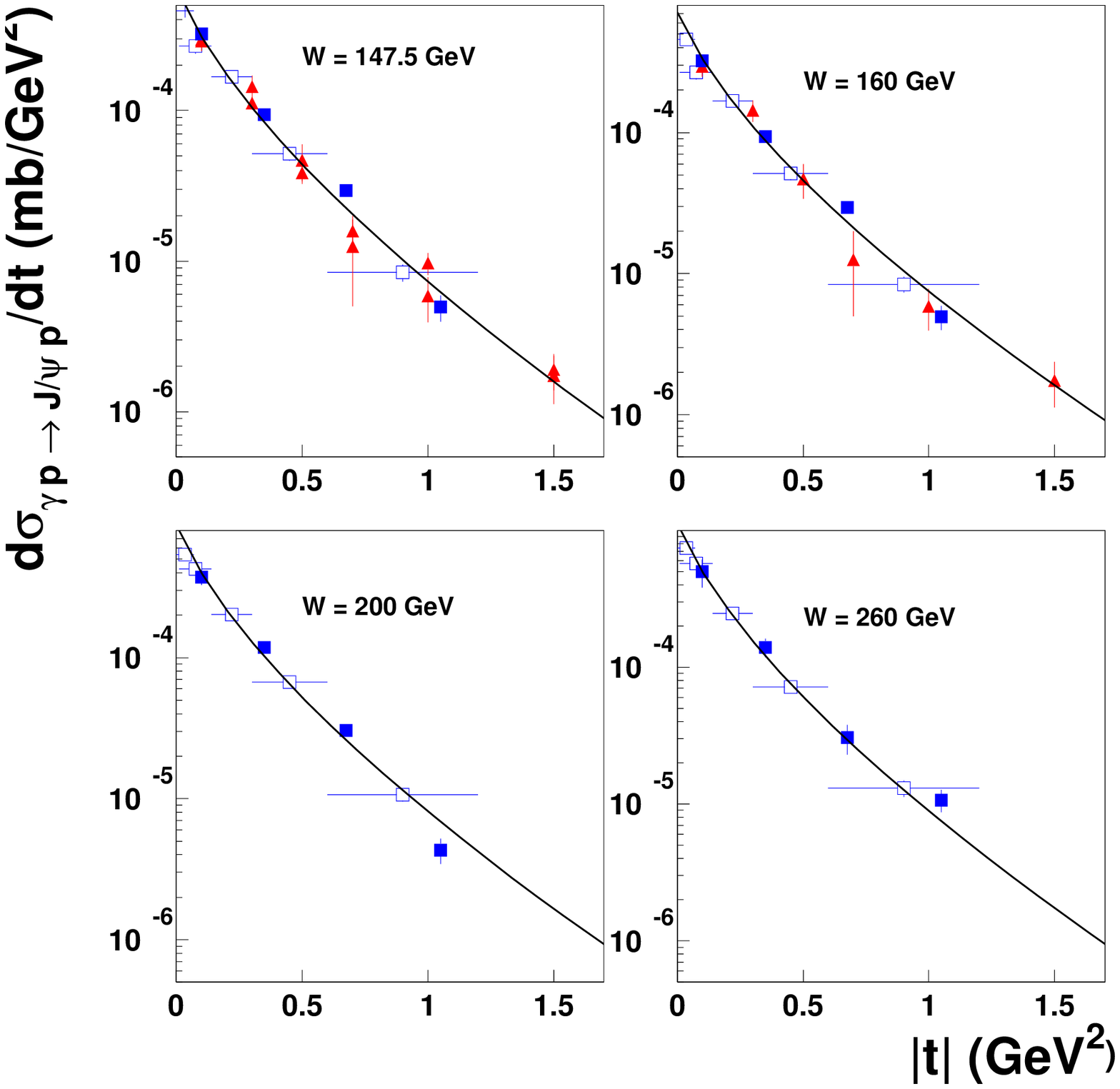}
\end{center}
\caption{\label{fig:jpsi_df1}  $J/\Psi$ differential cross
sections as a function of $t$ in the range of energy $W$ from 100
to 260 GeV.}
\end{figure}

The direct-channel exotic trajectory is (see \cite{Kononenko})
\beq
\alpha^E(s)=\alpha^E(0)+\alpha^E_1(\sqrt{s_0}-\sqrt{s_0-s})\,.
\eeq{exot} The relevant threshold value is
$s_0=(m_{J/\Psi}+m_P)^2$.

Let us remind also that $s,t$ and $u$ are not independent
variables but they are related by $s+t+u=\sum_{i} m_i^2\; = \; 2
m_{J/\Psi}^2 + 2m_P^2$

The integral of eqs. (\ref{VMD}), (\ref{dama}): \beq D(s,t,u)= c
(s-u) \int_0^1 {dz \biggl({1-z \over g}\biggr)^{-\alpha_t(t')}
\left( \biggl({z \over g} \biggr)^{-\alpha(s')-1}  -  \biggl({z
\over g} \biggr)^{-\alpha(u')-1} \right) }\,, \eeq{resulting} with
exotic trajectory (\ref{exot}) converges for any values of $s$ and
$t$ \footnote{Note that for $D(s,t)$ and $D(u,t)$, taken
separately, only the imaginary parts are convergent, while their
real parts diverge. However, for their difference, eq.
\ref{resulting}, the real part becomes also convergent.}. So, the
procedure of the analytic continuation introduced in Ref.
\cite{DAMA} and inevitable in the case of resonances, here can be
avoided, enabling numerical calculations with any desired
precision \footnote{We calculate the highly oscillating integral,
eq. \ref{resulting}, by using  the QUADPACK FORTRAN package
\cite{quadpack}.}.

\begin{table}[t]
\begin{center}
\begin{tabular}{|ll|ll|}
\hline
$\alpha^E(0)$ = & $-1.83$ & $\alpha^E_1(0)$ = & $0.01$ (GeV$^{-1}$) \\
$\alpha^P(0)$ = & $1.2313$ & $\alpha^P_1(0)$ = & $0.13498$ (GeV$^{-1}$)\\
$\alpha^P_2(0)$ = & $0.04$ (GeV$^{-2}$) & $t_2$ = & $36$ (GeV$^2$) \\
$g$ = & 13629 & $c$ = &  0.0025 \\
\hline
 & & $\chi^2/{d.o.f.}$ = & $0.83$ \\
\hline
\end{tabular}
\end{center}
\caption{Fitted values of the adjustable parameters
\label{fitpar}}
\end{table}

\section{Differential and Integrated Elastic Cross-Section} \label{s4}

To calculate the differential cross section we use the following
normalization: \beq {d\sigma\over{dt}}={1\over16\pi\lambda (s,
m_{J/\Psi}, m_P)}\vert D(s,t,u)\vert ^2\,, \eeq{dsigmadt} where
$\lambda (x,y,z) = x^2 + y^2 + z^2 - 2xy - 2yz - 2xz$.

The integrated elastic cross section is defined
as
\beq
\sigma_{el}(s)=\int_{-t_{max}=s/2}^{t_{thr.}\approx 0}dt\, {d\sigma\over{dt}}\,.
\eeq{sigmael}

\begin{figure}
\begin{center}
\includegraphics[width=0.45\textwidth,bb= 10 140 540 660]{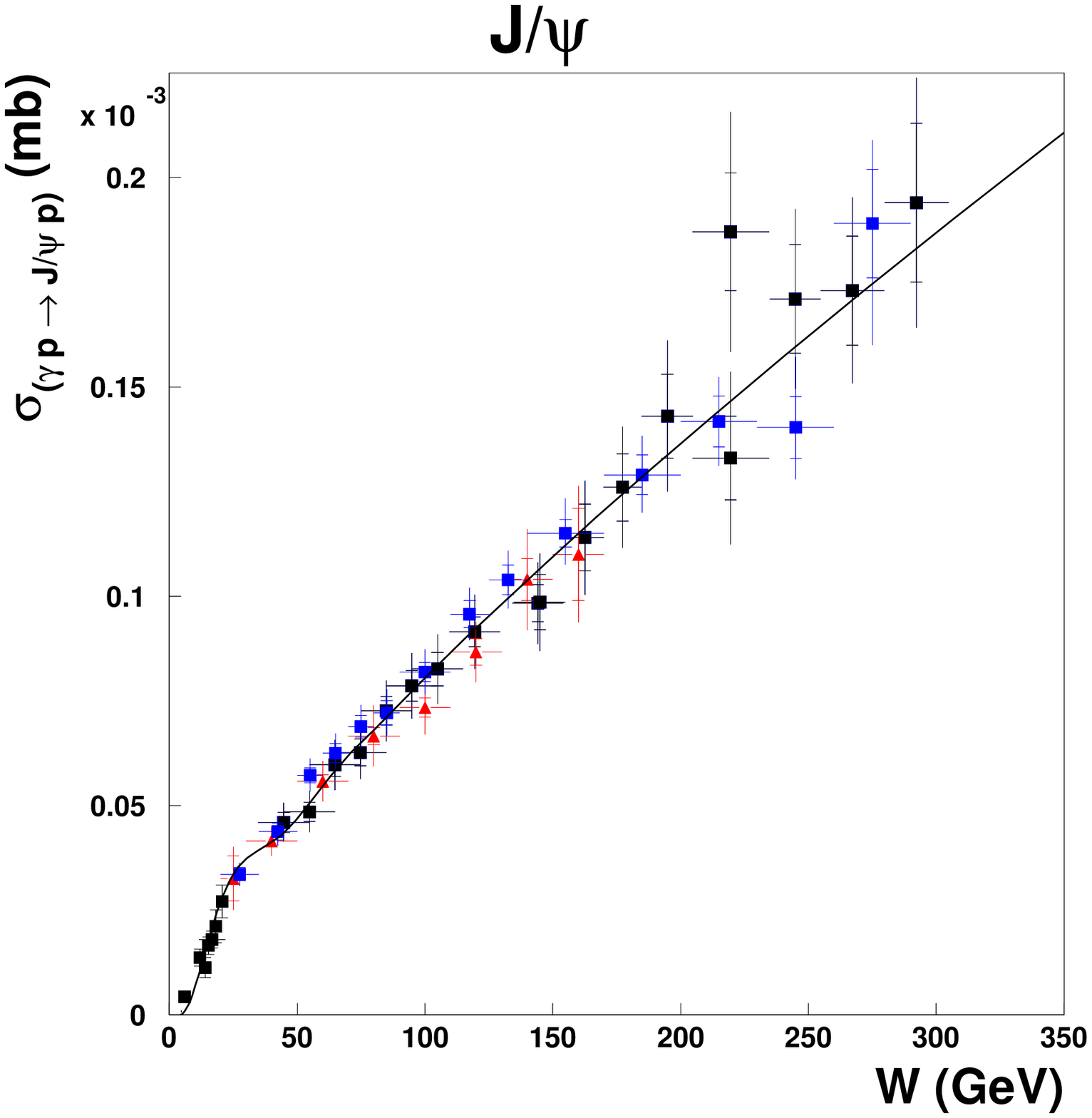}
\includegraphics[width=0.45\textwidth,bb= 10 140 540 660]{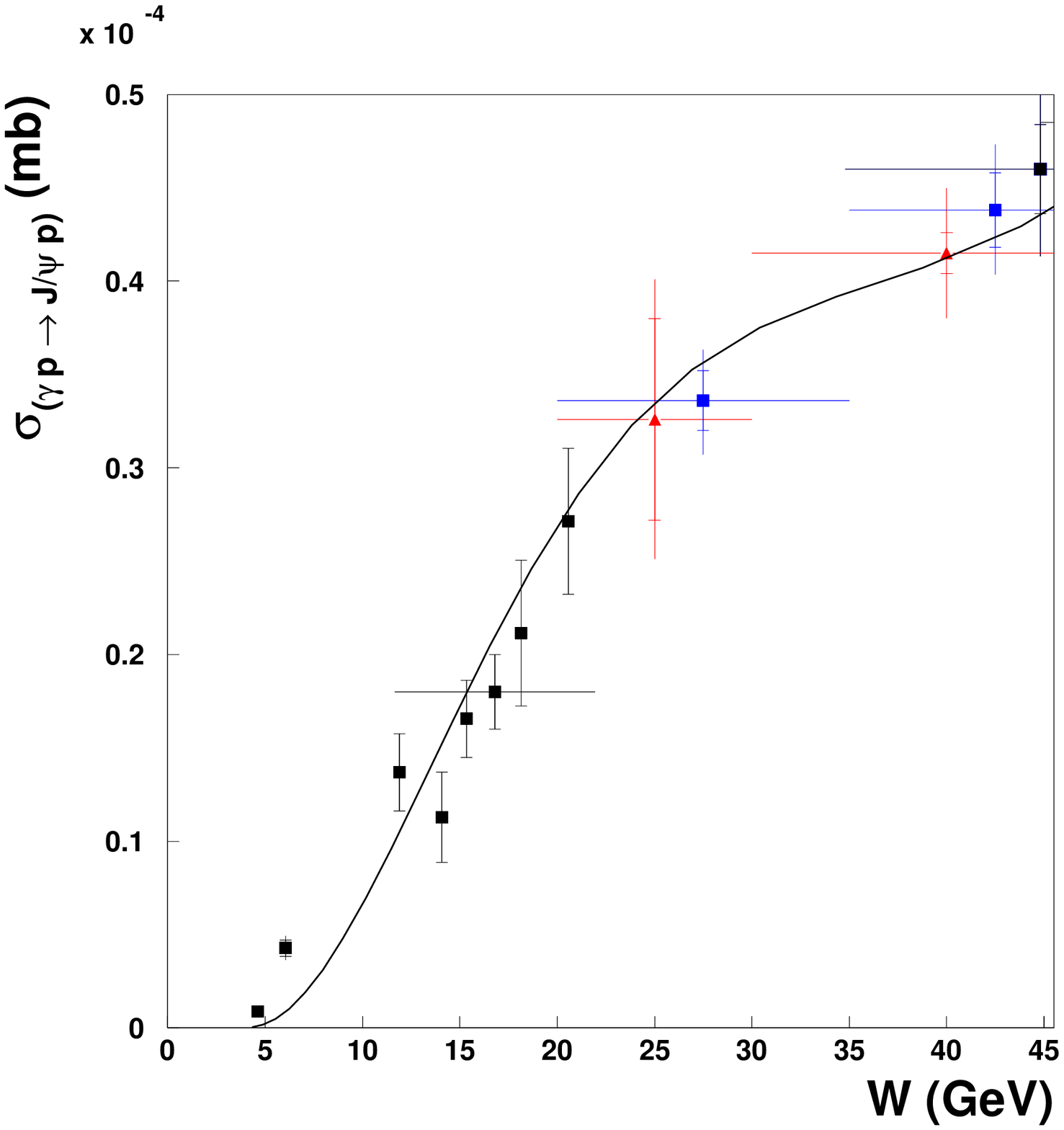}
\end{center}
\caption{\label{fig:jpsi_total}  $J/\Psi$ elastic cross section
for all energies (left panel) and close to the threshold region
(right panel).}
\end{figure}

We have fitted the parameters of the model to the data \cite{H1,
ZEUS} on $J/\Psi$ photoproduction differential cross section in
the energy range $W$ between 35 to 260 GeV. The results of the
fits together with the data are shown in Fig. \ref{fig:jpsi_df1}
and in the right panel of Fig. \ref{fig:jpsi_df}. The values of
the fitted parameters are quoted in Table \ref{fitpar}. With these
parameters we have plotted also the differential cross section at
lower energies (left panel of Fig. \ref{fig:jpsi_df}). It is
interesting to note that the shape of the cone (exponential
decrease in $t$), an important characteristics of diffraction,
survives at low energies,

From the fits to the differential cross section the integrated
elastic cross section was calculated, as shown in  Fig.
\ref{fig:jpsi_total}. The calculated curve is fully consistent
with the date in the whole kinematical region. Enlarged is shown
the energy region close to the threshold, where the background
contribution dominates.

The aim of this fit was to demonstrate the viability of the model
 covering the whole kinematical region - from the threshold to
highest energies and for all experimentally measured momenta
transfers. Yet, it includes the important region of ``low-energy"
diffraction, located between the threshold and Regge asymptotics
and described by a direct-channel exotic trajectory.

\section{Regge asymptotic behavior and the Pomeron trajectory}
\begin{figure}
\begin{center}
\includegraphics[width=8cm]{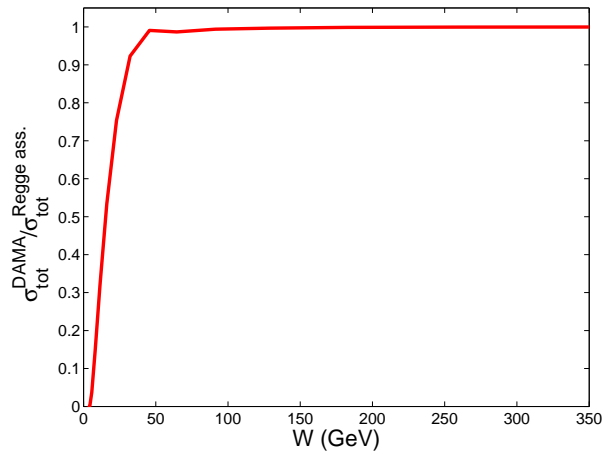}
\end{center}
\caption{\label{fig:10}  $J/\Psi$ total cross sections versus its
Regge asymptotics $\sigma_{tot}^{Regge. ass.}\sim
s^{\alpha^P(0)-1}$. Pure Regge asymptotic behaviour fails from
$50$ GeV downwards. }
\end{figure}

The high-energy behavior of the dual
amplitude is of Regge form, by definition, and we are very interested in the question
from which energy the Regge behavior starts to dominate. The
relevant energy can be extracted e.g. form  the total cross
section calculated from DAMA (with the parameters fitted to the
data) divided by its asymptotic form $s^{\alpha^P(0)-1},$ as shown
in Fig. \ref{fig:10}. One can see that the ratio starts to
deviate from $1$ around $50$ GeV, which means that below this energy down to the threshold
non-asymptotic, non-Regge effects become important, the contribution of the background is not anymore negligible.

Beyond $50$ GeV DAMA is typically Regge behaved. The remaining
detail affecting the Regge behavior are: the form of the Regge
singularity (here, a simple Regge pole) and the form of the
Pomeron trajectory.

\begin{figure}
\begin{center}
\includegraphics[width=0.45\textwidth,bb= 10 140 540 660]{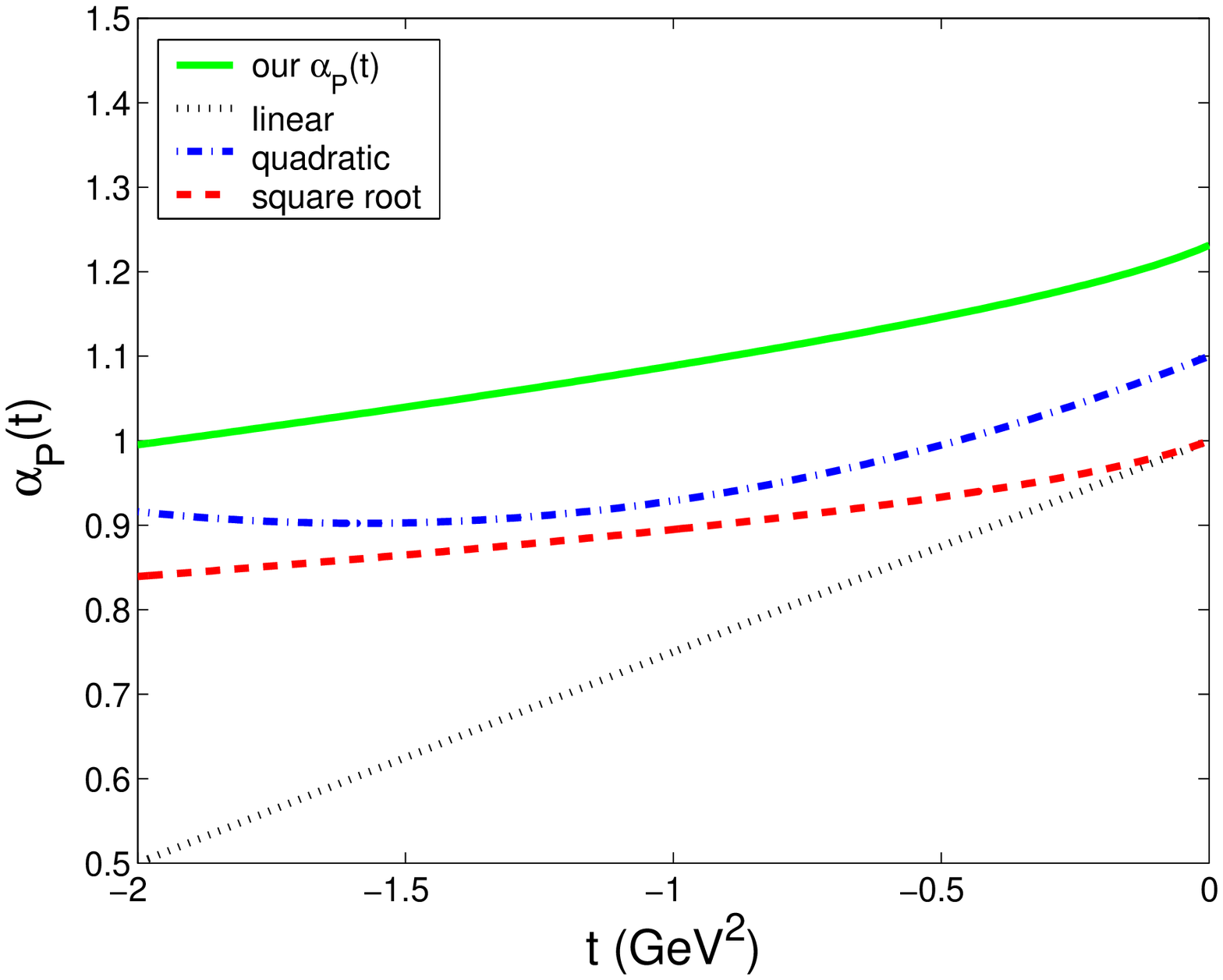}
\includegraphics[width=0.45\textwidth,bb= 10 140 540 660]{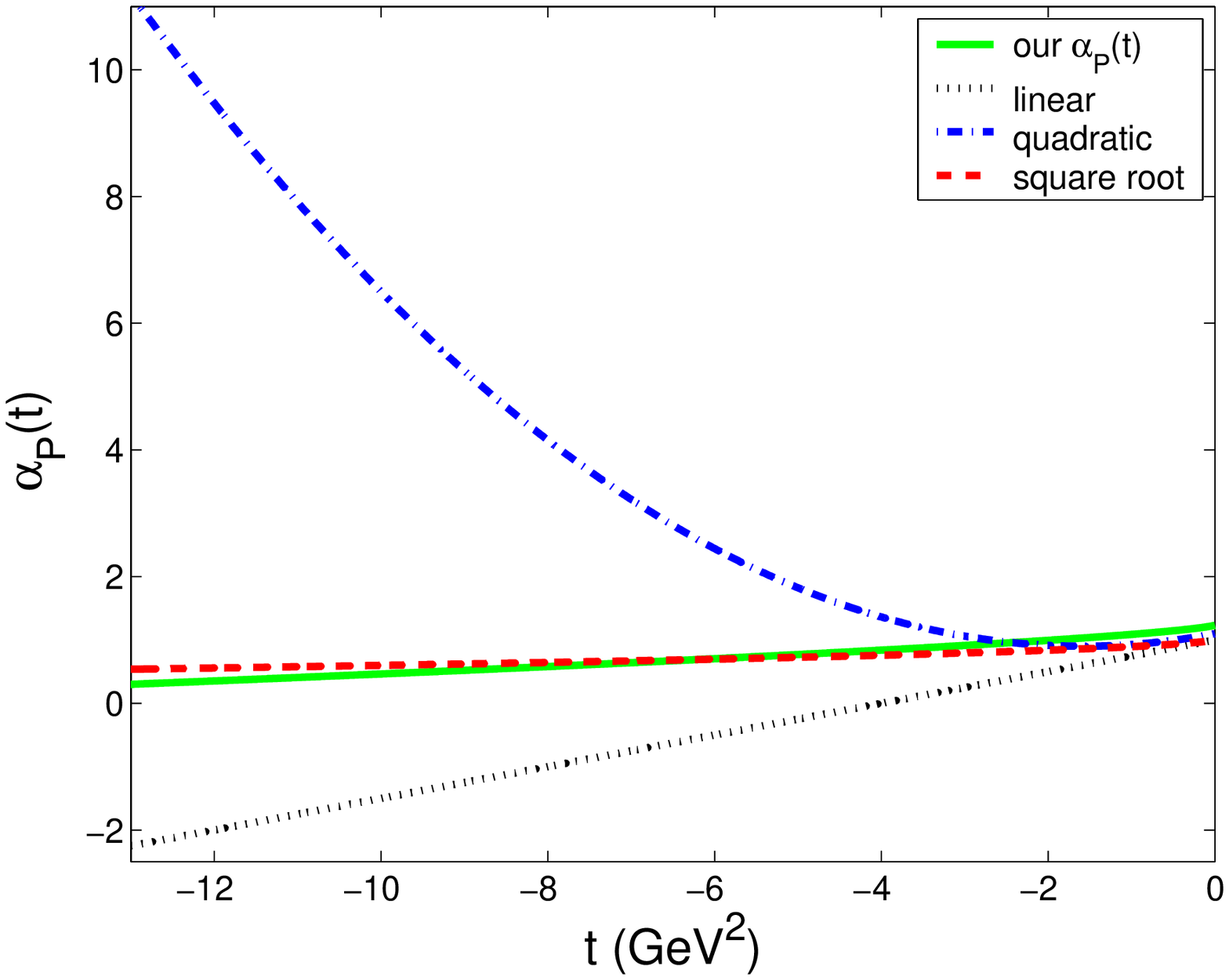}
\end{center}
\caption{\label{fig:Pomeron} Pomeron trajectories used in Refs.
\cite{Brandt, FJPP, ProkMart}, see text.}
\end{figure}

In the present paper, to fit the rise of the cross sections with
energy, we used a "supercritical" Pomeron trajectory, with
$\alpha(0)=1.2313$ (see Table I). Alternatively, in a dipole
Pomeron model \cite{FJPP, ProkMart} cross sections rise with
energy logarithmically. In other words,
various versions and fits of Regge-type models give different
Pomeron trajectories, as shown in Fig. \ref{fig:Pomeron}.

Fig. 7 shows, apart from our trajectory (\ref{Pomer}) with
the parameters presented in Table I, \\
1) the simplest linear
trajectory \cite{FJPP}a), denoted in the Fig. \ref{fig:Pomeron} as
"linear": 
\beq \alpha^P(t)=1.0 + 0.25\cdot {\rm GeV}^{-2}\cdot
t\,; 
\eeq{linear} 
2) the trajectory of Ref. \cite{Brandt}
containing a quadratic term (denoted "quadratic"): 
\beq
\alpha^P(t)=1.1 + 0.25\cdot {\rm GeV}^{-2}\cdot t + 0.078\cdot
{\rm GeV}^{-4}\cdot t^2 \,; 
\eeq{quad} 
3) the trajectory with
one square root threshold \cite{FJPP}b), denoted as "square root":
\beq \alpha^P(t)=1.0+0.138\cdot {\rm GeV}^{-1}\cdot
(\sqrt{t_1}-\sqrt{t_1-t})\,, \quad t_1=4m_\pi^2\,. \eeq{sqrt}

Although, as can be seen in Fig. 7, the intercepts of these
trajectories are quite different (depending on the type of the
Pomeron singularity), their slopes at small $|t|$ are nearly the
same. The quadratic term in the Pomeron trajectory of
\cite{Brandt} (chain lines in Fig. \ref{fig:Pomeron}) rises
dramatically, thus limiting its applicability to small values of
$|t|$ (actually, it was fitted to the data at $|t|\le 2$ GeV).

The large-$|t|$ behavior of the nonlinear trajectories is an
interesting problem by itself. As is well known, wide-angle
scaling behavior of the amplitude requires the a logarithmic
asymptotic behavior of the trajectory. More details on this
interesting subject can be found in a recent paper \cite{Capua}
and references therein.

\section{Conclusion} \label{s6}

The low-energy part of the present model can serve as a background
both in theoretical and experimental studies. We know from
high-energy diffraction (=Pomeron) that it is universal. From the
duality arguments one would assume that the same is true for the
background, dual to the Pomeron. However, the present model of the
background contains a reaction-dependent parameter, namely the
value of the direct-channel threshold mass violating this
universality. To see the role of the background separately, more
fits to other reactions will be necessary.

In any case, the model presented in this paper offers a
complementary approach to soft dynamics of strong interactions,
namely to its component dominated by diffraction, which is beyond
the scope of the perturbative quantum chromodynamics.

An important finding of the present paper is the behavior of the
differential cross sections at low energies, calculated with the
parameters fitted to high-energy data, and presented in the left
panel of Fig. 3. It shows the existence of a shrinking forward
cone, typical of diffraction, persisting to lowest energies,
dominated by the background (cf. Fig. 6).

The proper parameterization of the background is important for most
of the reactions, such as $pp,\ \ p\bar p$ or $\pi p$ scattering, and 
thus the obtained results are universal, and the calculated background
can and should be included in any model or data analyses at low
energies.

In this paper we have used the simples version of DAMA. The model
can be extended to include, for example, a dipole Pomeron, which 
can be generated by differentiating eq. (1) in $\alpha(t)$, as was
done, e.g., in Ref. \cite{FJPP}.

Generalizations to non-zero $Q^2$ (electroproduction) of
Regge-pole models can be found in Refs. \cite{FJPP, Capua} and
\cite{ProkMart}, and those of DAMA were studied in \cite{JMP, MDAMA}. It
should be however noted that at high $Q^2$ vector meson dominance
may not be valid anymore. The ambitious program of a unified
description of soft and hard dynamics, however, is beyond the
scope of the present paper.

 \vskip
0.2cm We thank A. Bugrij, A. Papa and E. Predazzi for the fruitful discussions
and Yu. Stelmakh for his help in preparing the manuscript. This 
work was supported in part by the Ministero Italiano dell'Universita' e
della Ricerca.


\end{document}